\documentclass[sigconf]{acmart}

\AtBeginDocument{%
 }

\usepackage[most]{tcolorbox}
\usepackage{xcolor}
\usepackage{booktabs}

\definecolor{themeRed}{HTML}{F25050}
\definecolor{themeBlue}{HTML}{506AF2}

\tcbset{
  highlights/.style={
    enhanced,
    colframe=blue!30,
    colback=blue!60,
    boxrule=0pt,
    corners=south,
    rounded corners=north,
    arc=1mm,
    boxsep=0pt,
    left=1mm,
    right=1mm,
    top=0.5mm,
    bottom=0.5mm, fontupper=\bfseries\color{white},
  },
}
\newtcolorbox{custombox}[1]{
	colback=gray!10,
	colframe=gray!70,
	left=1mm,
	right=1mm,
	top=1mm,
	bottom=1mm,
	fonttitle=\bfseries,
	arc=0mm,
	leftrule=1mm,
	rightrule=0mm,
	toprule=0mm,
	bottomrule=0mm,
	notitle,
	before=\par\smallskip\noindent,
	before upper={\textbf{#1: } },
}

\copyrightyear{2025}
\acmYear{2025}
\setcopyright{acmlicensed}\acmConference[ITiCSE 2025]{Proceedings of the 30th ACM Conference on Innovation and Technology in Computer Science Education V. 1}{June 27-July 2, 2025}{Nijmegen, Netherlands}
\acmBooktitle{Proceedings of the 30th ACM Conference on Innovation and Technology in Computer Science Education V. 1 (ITiCSE 2025), June 27-July 2, 2025, Nijmegen, Netherlands}
\acmDOI{10.1145/3724363.3729088}
\acmISBN{979-8-4007-1567-9/2025/06}

\begin{document}


\title[Neurodiversity in Computing Education Research: A Systematic Literature Review]{Neurodiversity in Computing Education Research: A Systematic Literature Review}


\author{Cynthia Zastudil}
\affiliation{
    \institution{Temple University}
    \city{Philadelphia}
    \state{PA}
    \country{United States}
}
\email{cynthia.zastudil@temple.edu}

\author{David H. Smith IV}
\affiliation{
    \institution{University of Illinois}
    \city{Urbana}
    \state{IL}
    \country{United States}
}
\email{dhsmith2@illinois.edu}

\author{Yusef Tohamy}
\affiliation{
    \institution{Temple University}
    \city{Philadelphia}
    \state{PA}
    \country{United States}
}
\email{yusef.tohamy@temple.edu}

\author{Rayhona Nasimova}
\affiliation{
    \institution{Temple University}
    \city{Philadelphia}
    \state{PA}
    \country{United States}
}
\email{rayhana.nasimova@temple.edu}

\author{Gavin Montross}
\affiliation{
    \institution{Temple University}
    \city{Philadelphia}
    \state{PA}
    \country{United States}
}
\email{gavin.montross@temple.edu}

\author{Stephen MacNeil}
\affiliation{
    \institution{Temple University}
    \city{Philadelphia}
    \state{PA}
    \country{United States}
}
\email{stephen.macneil@temple.edu}

\renewcommand{\shortauthors}{Cynthia Zastudil et al.}

\begin{abstract}
Ensuring equitable access to computing education for all students---including those with autism, dyslexia, or ADHD---is essential to developing a diverse and inclusive workforce. To understand the state of disability research in computing education, 
we conducted a systematic literature review of research on neurodiversity in computing education. 
Our search resulted in 1,943 total papers, which we filtered to 14 papers based on our inclusion criteria.
Our mixed-methods approach analyzed research methods, participants, contribution types, and findings.  The three main contribution types included empirical contributions based on user studies (57.1\%), opinion contributions and position papers (50\%), and survey contributions (21.4\%). Interviews were the most common methodology (75\% of empirical contributions). There were often inconsistencies in how research methods were described (e.g., number of participants and interview and survey materials). Our work shows that research on neurodivergence in computing education is still very preliminary. Most papers provided curricular recommendations that lacked empirical evidence to support those recommendations. Three areas of future work include investigating the impacts of active learning, increasing awareness and knowledge about neurodiverse students' experiences, and engaging neurodivergent students in the design of pedagogical materials and computing education research.



%

\end{abstract}





\maketitle

\section{Introduction}


Over the last decade, computing education (CE) researchers are increasingly embracing the goal of broadening access to computing for many underrepresented groups. These efforts include addressing gender~\cite{ michell2017broadening, menier2021broadening} and racial~\cite{scott2017broadening, london2020systematic, cole2008examining} disparities. 
However, there is still a need to better understand the experiences of diverse student populations~\cite{blaser2020why}.
Disabled students\footnote{In this paper, we primarily use identity first language (e.g., ``autistic students''), as recent research has found that many individuals prefer identity first language~\cite{sharif2022should}, but we recognize that some people also prefer person-first language (e.g., ``students with autism'') and their preferences should be respected.}, especially those with invisible disabilities, can face educational barriers, including a lack of appropriate accommodations~\cite{borsotti2024neurodiversity, shinohara2020access}, limited awareness of their disabilities~\cite{borsotti2024neurodiversity}, and difficulties developing a sense of belonging~\cite{runa2023student}.

Disability is highly personal~\cite{lutz2005disability}; 
the access needs of physically disabled students may differ significantly from the needs of neurodivergent students. 
When we use the term `neurodivergent students', we are referring to students whose cognitive processing styles differ from socially or culturally constructed norms, which often results in unique strengths and challenges. Neurodivergent refers to a wide variety of identities, including autism, dyslexia, ADHD, and others. 

Researchers have emphasized the need for research about the experiences of neurodivergent learners in CE~\cite{luchs2021considering, koushik2019broadens}. 
While there has been an effort to increase awareness of different access needs of diverse student populations and help students develop empathy~\cite{zhao2020comparison, cotler2024enhancing}, it is also important to ensure that CE curricula and pedagogical methods are accessible to more than just neurotypical students.
In 2021, researchers conducted a comprehensive systematic literature review in which they examined the trends of accessibility research from 1994 to 2019~\cite{mack2021we}. While their review provided valuable general insights about accessibility, it did not address the specific challenges and opportunities within CE. There remains an urgent need to better understand neurodiversity in CE to design curricula and pedagogies that better serve more students.
Recently, there has also been an increasing focus on the accessibility of K-12 CE~\cite{israel2015empowering, blaser2024accessibility}. However, insights about K-12 students may not apply to neurodiverse computing students at the post-secondary level. 
For example, educators tend to assume that post-secondary students are more independent than their K-12 counterparts~\cite{van2015higher} and therefore better equipped to self-advocate~\cite{santhanam2024comparison} or manage their schedules~\cite{van2015higher}.
We expand on this prior research by focusing on research about neurodiversity in post-secondary computing education. We investigate the following research questions: 
\begin{itemize}
    \item[\textbf{RQ1:}] What research methodologies and contributions have researchers used to study neurodiversity in CE research?
    \item[\textbf{RQ2:}] What findings and outcomes have emerged from prior CE research on neurodiversity in CE?
\end{itemize}

Our search of the literature resulted in 1,943 papers, which we further filtered to 14 papers based on our inclusion criteria. 
Our analysis of the papers revealed three main contribution types: user studies (57.1\%), position papers (50.0\%), and literature reviews (21.4\%). Interviews were the most common methodology (75\% of user studies). However, there were often inconsistencies in reporting details on the user studies (e.g., number of participants and interview and survey materials). The position papers all consisted of recommendations for inclusive curriculum design. Our work shows that research on neurodivergence in CE is still very preliminary. A large portion of papers consisted of curricular recommendations; however, many of these were position papers and did not have empirical evidence underpinning the recommendations. We believe that future work has the opportunity to investigate the impacts of active learning and how to increase awareness about neurodiverse students' access needs. We also highlight the need to engage neurodiverse students in the design and research process.


\section{Methodology}
To understand how neurodiversity has been studied in CE research, we conducted a systematic literature review. 
Below, we discuss our exclusion and inclusion criteria, search and selection processes, and analysis strategies.

\subsection{Exclusion \& Inclusion Criteria}
\label{sec:criteria}
The goal of our work is to investigate the research methods, contributions, goals, and findings of CE research focused primarily on neurodiversity in post-secondary CE. We used exclusion and inclusion criteria (shown in Table~\ref{tab:excl-incl}) which limited our results to a representative but reasonable scope. 

\begin{table}[]
\small
\centering
\begin{tabular}{@{}l@{}}
\toprule
\textbf{Exclusion Criteria} \\ \midrule
\textbf{EC$_1$:} Not an academic paper \\
\textbf{EC$_2$:} Not available in English \\
\toprule
\textbf{Inclusion Criteria} \\ \midrule
\begin{tabular}[c]{@{}l@{}}\textbf{IC$_1$:} About neurodiversity (e.g., autism spectrum disorder, ADHD, dyslexia)\end{tabular} \\
\textbf{IC$_2$:} About computing education (e.g., computing students or courses) \\
\begin{tabular}[c]{@{}l@{}}\textbf{IC$_3$:} About post-secondary computing education\end{tabular} \\ \bottomrule
\end{tabular}
\caption{Exclusion and inclusion criteria.}
\label{tab:excl-incl}
\end{table}

\subsection{Search \& Selection Process}
\label{sec:search}

We conducted our selection process according to best practices in Preferred Reporting Items for
Systematic Reviews and Meta-Analysis (PRISMA) reporting~\cite{page2021prisma} in three phases. \textbf{Identification Phase} - searching the selected databases to identify potentially eligible references. \textbf{Screening Phase} -  reviewing the resulting set of references from the identification phase to remove references that meet the exclusion criteria. \textbf{Eligibility Phase} - reviewing the resulting set of papers from the screening phase to ensure that included papers meet the inclusion criteria. These three phases are visualized in more detail in Figure~\ref{fig:prisma}.

\subsubsection{Identification Phase}
To ensure a broad and representative sample, we used three databases in our literature review: ACM Digital Library, IEEE Xplore, and Scopus. The ACM Digital Library and IEEE Xplore databases include publications from many conference venues that are highly influential in CE research (e.g., SIGCSE Technical Symposium, ITiCSE, and FIE)~\cite{apiola2023past} and Scopus includes many academic journals in relevant research areas.
Our search query was developed by conducting a preliminary search of the literature to determine keywords that are commonly used in studies of neurodiversity in CE research. By looking at co-occurring keywords in papers identified in our preliminary search, we constructed an initial set of keywords. We further expanded this set by including related terms, resulting in the following search query: 

\noindent \texttt{(``computer science education''
    OR ``cs education'' OR ``computing education'' OR ``inclusive education'') AND (``cognitive disability'' OR ``cognitive impairment'' OR ``cognitive accessibility'' OR ``autism'' OR ``asd'' OR dys* OR neurodiver* OR ``intellectual disability'' OR ``learning disability'' OR ``adhd'')}

The search results were collected in January 2025 (ACM DL and IEEE Xplore) and March 2025 (Scopus), and represent all papers published up to that point in time that matched our search terms. We identified 1,943 references from the ACM Digitial Library (n = 1,692), IEEE Xplore (n = 120), and Scopus (n = 131).

\subsubsection{Screening Phase}
We identified and removed duplicates, references that were not full papers (e.g., conference proceedings, workshops, and posters), and references not available in English. We removed 64 duplicate references, 402 references which were not full papers, and 1 reference not available in English. After screening, we were left with 1,476 references to review for eligibility.

\begin{figure*}
    \centering
    \includegraphics[width=0.9\linewidth]{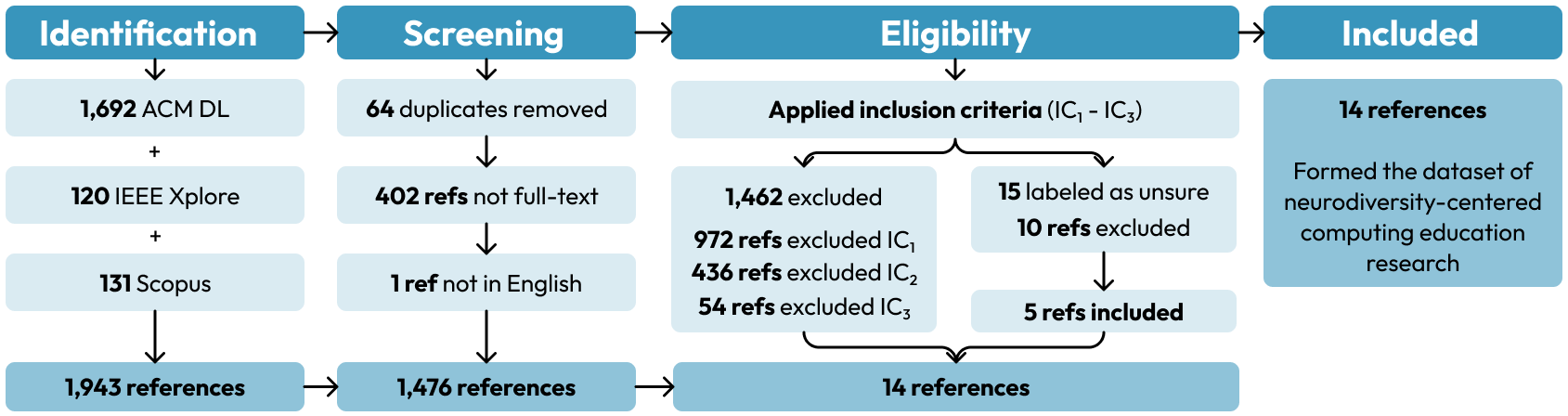}
    \caption{PRISMA diagram detailing our identification, screening, and assessment of eligibility of papers for inclusion.}
    \label{fig:prisma}
\end{figure*}

\subsubsection{Eligibility Phase}
During the eligibility phase, we identified the papers which met our inclusion criteria (see Table~\ref{tab:excl-incl}). This phase included two members of our research team reviewing the title and abstracts of every paper in the remaining set of 1,476 papers to determine whether or not it met the inclusion criteria and marking each paper as include, exclude, or unsure. For each paper that was excluded, a reason was provided. 
Each person's reviews were kept confidential until all papers had been reviewed. Once all of the papers had been reviewed for inclusion, any disagreements were mitigated via discussion by the reviewing researchers. Any papers marked unsure by any of the reviewing researchers were discussed to determine whether or not it should be included by reviewing the papers' details and applying the inclusion and exclusion criteria together to reach a consensus. We always reached a consensus. In total, 1,462 papers were excluded and 15 were marked unsure. Of the 15 papers marked unsure, 5 were included and 10 were excluded. For detailed information about the reasons why papers were excluded, see Figure~\ref{fig:prisma}. A total of 14 papers were included in this study~\cite{egan2005students, dixon2007comparative, martinez2011mhy, pilotte2016autism, assiter2018experiences, sitbon2018engaging, stuurman2019autism, ross2019supporting, begel2021remote, kletenik2024awareness, kirdani2024neurodivergent, haynes2024neurodiverse, borsotti2024neurodiversity, sharmin2024towards}.

\subsection{Data Analysis}

Two members of the research team analyzed the resulting set of 14 papers.  To investigate RQ1, we applied a deductive coding scheme based on a systematic literature review of accessibility research by Mack et al.~\cite{mack2021we}.
In addition to the modified codes inspired by Mack et al., we included codes for the pedagogical approaches (e.g., active learning methodologies and teaching strategies) used, as done in previous reviews~\cite{shehzad2023literature, loraas2021study}. 
We included all the codes from Mack et al.~\cite{mack2021we}; however, many of the codes required modification due to the subject-specific nature of our review. 
Codes that were deemed irrelevant to the current context were removed, while new codes were added to better capture elements pertinent to computing education. 
Lastly, we included the proportion of people with disabilities included in the study in cases where the study participants were of mixed-ability.
Each included paper was read by two members of the research team and the codes were applied. If there were disagreements, the coding researchers discussed their reasoning for applying the code to reach a consensus. 

\begin{table*}[]
\centering
\small
\begin{tabular}{@{}llc@{}}
\toprule
\textbf{Category} & \textbf{Codes} & \textbf{Multiple?} \\ \midrule
Community of Focus & Autism Spectrum Disorder,  ADHD,  Dyslexia/Dyspraxia/Dyscalculia, Learning Disability, Other& Yes \\
Study Method & \begin{tabular}[t]{@{}l@{}}Controlled experiment, Usability testing, Field studies, Interviews, Questionnaires, \\ Case studies, Focus groups, Workshop or design session, Randomized control trials, Other\end{tabular} & Yes \\
Participatory Design Use & Yes, No & No \\
User Study Location & Classroom study, Research lab study,  Online/remote, Other & Yes \\
Participant Groups & Neurodivergent people, Non-Neurodivergent people, No user study & Yes \\
Use of Proxies & Yes, No & No \\
Ability-Based Comparisons & Yes, No & No \\
Contribution Type & Empirical, Artifact, Survey, Methodological,  Theoretical and opinion, Dataset & Yes \\\bottomrule
\end{tabular}
\caption{The codes used in analysis of the 14 papers selected. Multiple refers to whether or not multiple codes may apply.}
\label{tab:codes}
\end{table*}

To investigate RQ2, we conducted a thematic analysis of the findings and discussion sections from our included papers. Our thematic analysis focused primarily on methods, contribution types, and the primary findings of each paper. Two members of the research team created descriptions of the course(s) in which the research was conducted, the goals of the research, challenges and opportunities, and the primary findings. The first author then reviewed these descriptions and identified themes.

\section{Results}

\begin{figure}
    \centering
    \includegraphics[width=0.9\linewidth]{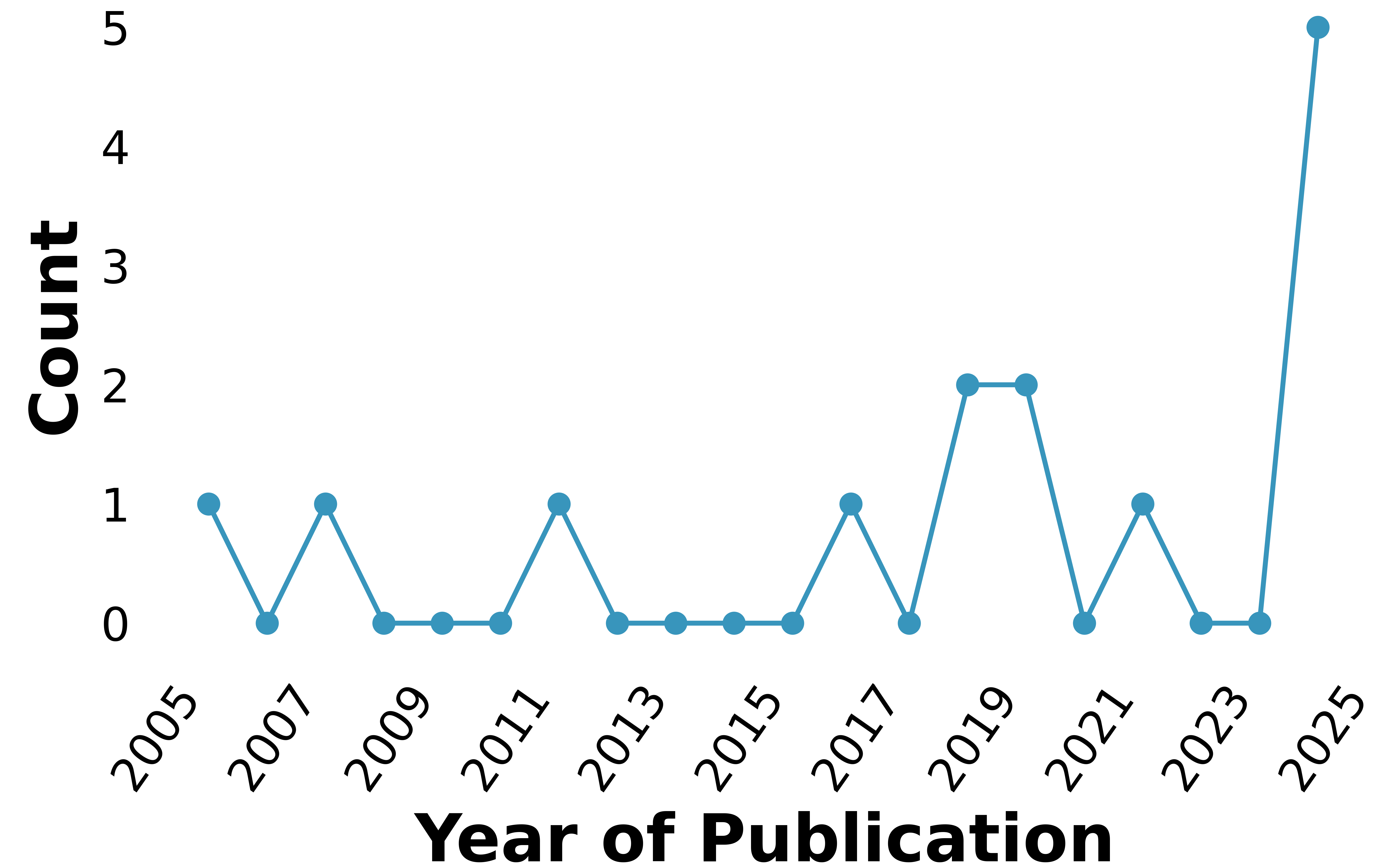}
    \caption{Paper counts by year of publication.}
    \label{fig:year-of-pub}
\end{figure}

The number of papers researching neurodiversity in CE has been steadily increasing, from 2 in 2018 to 5 in 2024, as shown in Figure~\ref{fig:year-of-pub}. In Section~\ref{sec:rq1}, we present the results of our quantitative coding of these papers' methodologies and contributions to address RQ1. In Section~\ref{sec:rq2}, we present the results of our thematic analysis of these papers' findings and outcomes to address RQ2.

\subsection{RQ1: Methods and Contribution Types}
\label{sec:rq1}



\subsubsection{Research Methods}

The most common methodologies used were argumentation, which was typically used in position papers~\cite{stuurman2019autism, pilotte2016autism, kletenik2024awareness, martinez2011mhy, egan2005students, rose2013universal, sharmin2024towards}, and interviews~\cite{dixon2007comparative, kletenik2024awareness, begel2021remote, haynes2024neurodiverse, borsotti2024neurodiversity, kirdani2024neurodivergent}. 
Other frequently used methodologies were literature reviews~\cite{pilotte2016autism, stuurman2019autism, sharmin2024towards}, questionnaires~\cite{kletenik2024awareness, begel2021remote, haynes2024neurodiverse}, case studies~\cite{sitbon2018engaging, haynes2024neurodiverse}, experience reports~\cite{assiter2018experiences, begel2021remote}, participatory design~\cite{sitbon2018engaging}, and usability testing~\cite{dixon2007comparative}.

Of the literature reviews, 2 were narrative reviews~\cite{pilotte2016autism, stuurman2019autism}. The remaining literature review was a systematic literature review~\cite{sharmin2024towards}. We chose to include this systematic literature review on its own, rather than to include all of the papers included in their review as they broadly included STEM-related papers in their search and did not provide a full list of the papers included.

Of the papers selected, 8 of them contained user studies. These user studies were frequently conducted online~\cite{begel2021remote, haynes2024neurodiverse, kirdani2024neurodivergent, borsotti2024neurodiversity}. However, studies also took place in-person~\cite{borsotti2024neurodiversity}, in classrooms~\cite{assiter2018experiences}, and within a disability support organization~\cite{sitbon2018engaging}. Two papers did not explicitly specify where the user study was conducted~\cite{dixon2007comparative, kletenik2024awareness}. In papers that featured a user study, only half of them (n = 4) included the study materials, such as interview scripts, surveys, course materials, or external materials (e.g., textbooks)~\cite{begel2021remote, haynes2024neurodiverse, kirdani2024neurodivergent, assiter2018experiences}.

In only one case~\cite{dixon2007comparative} did authors make an ability-based comparison (i.e., explicitly comparing the performance or experiences of neurodivergent and non-neurodivergent students). Papers rarely used proxies (i.e., a non-disabled person provides their thoughts as someone who would be affected by a solution as if they were disabled~\cite{andresen2001proxy}) in their studies~\cite{sitbon2018engaging, kletenik2024awareness}.

\begin{figure*}
    \centering
    \includegraphics[width=.85\linewidth]{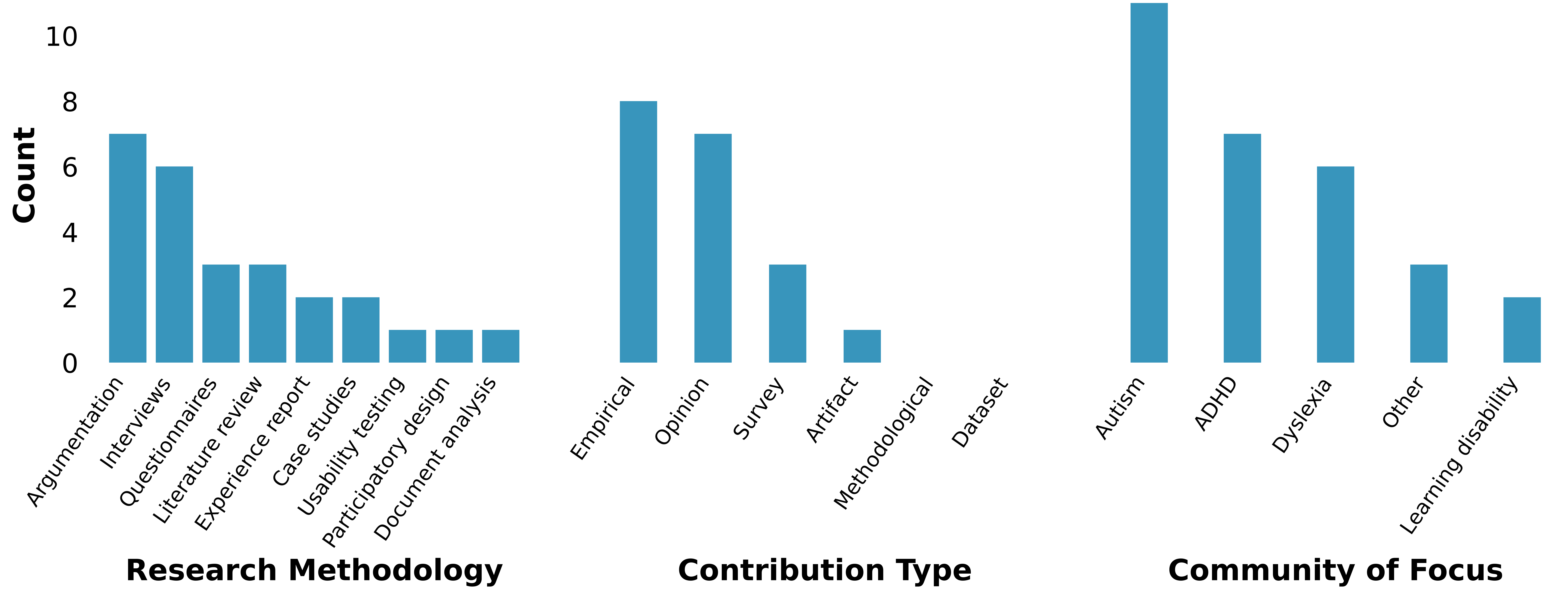}
    \caption{The research methodologies~\cite{mack2021we}, contribution types~\cite{wobbrock2016research}, and communities of focus of the papers~\cite{mack2021we} we analyzed. Papers can contain more than one type of research methodology, contribution type, and community of focus.}
    \label{fig:results}
\end{figure*}

\subsubsection{Contribution Types}

We used Wobbrock's taxonomy for research contributions~\cite{wobbrock2016research} (contributions are not mutually exclusive) to better understand the types of knowledge contributed to research on neurodivergence in CE. We found that the majority (n = 8) of the papers made empirical contributions~\cite{dixon2007comparative, sitbon2018engaging, assiter2018experiences, kletenik2024awareness, begel2021remote, haynes2024neurodiverse, borsotti2024neurodiversity, kirdani2024neurodivergent}, followed by opinion contributions (n = 7)~\cite{pilotte2016autism, stuurman2019autism, kletenik2024awareness, martinez2011mhy, egan2005students, ross2019supporting, sharmin2024towards} in the form of position papers. There were also 3 surveys (i.e., literature review) contributions~\cite{pilotte2016autism, stuurman2019autism, sharmin2024towards}. These survey contributions were accompanied by opinion contributions. 
One paper included an artifact contribution by sharing their teaching materials~\cite{begel2021remote}.

\subsubsection{Pedagogical Techniques Studied}

Authors evaluated pedagogical techniques and approaches in 5 papers with user studies. However, only 2 of these papers directly studied the impacts of the pedagogy used~\cite{dixon2007comparative, haynes2024neurodiverse}. Dixon's paper evaluated the usability of an algorithm visualization software for dyslexic and non-dyslexic students. Haynes-Magyar's paper presented case studies assessing the accessibility of Parsons problems~\cite{ericson2022parsons} for students with different conditions. The other 3 papers detailed the use of their pedagogical approaches as part of an experience report~\cite{begel2021remote, ross2019supporting, assiter2018experiences, borsotti2024neurodiversity}.

\subsubsection{Community of Focus \& User Study Participants}

Half of the papers we analyzed (n = 7) included one community of focus. The other half of the papers included between 2--7 types of neurodivergence.
The most common community of focus was autism spectrum disorder~\cite{pilotte2016autism, stuurman2019autism, assiter2018experiences, kletenik2024awareness, begel2021remote, haynes2024neurodiverse, borsotti2024neurodiversity, egan2005students, ross2019supporting, sharmin2024towards, kirdani2024neurodivergent}\footnote{In 2013, Asperger's Syndrome was removed from the DSM-5 as an official diagnosis~\cite{regier2013dsm}, and it is now considered to be a part of the autism spectrum. For the purposes of this paper, we consider Egan's work~\cite{egan2005students} to be focused on autistic students.}, followed by ADHD~\cite{assiter2018experiences, kletenik2024awareness, haynes2024neurodiverse, borsotti2024neurodiversity, ross2019supporting, sharmin2024towards, kirdani2024neurodivergent}, dyslexia, dyspraxia, or dyscalculia~\cite{dixon2007comparative, assiter2018experiences, kletenik2024awareness, martinez2011mhy, borsotti2024neurodiversity, ross2019supporting}, and learning disabilities~\cite{assiter2018experiences, haynes2024neurodiverse}. There were 3 papers which fell into the ``Other'' category. This included intellectual disability~\cite{sitbon2018engaging}, obsessive-compulsive disorder~\cite{kirdani2024neurodivergent}, persistent post-concussion syndrome~\cite{borsotti2024neurodiversity}, fibromyalgia~\cite{borsotti2024neurodiversity}, Tourette's syndrome~\cite{haynes2024neurodiverse}, a seizure disorder~\cite{haynes2024neurodiverse}, complex post-traumatic stress disorder~\cite{borsotti2024neurodiversity}, cyclothomia~\cite{borsotti2024neurodiversity}, and memory impairments~\cite{haynes2024neurodiverse}.

In the case of user studies, the median number of participants was 8, with two outliers: Kirdani-Ryan' and Ko's work had 21 participants~\cite{kirdani2024neurodivergent} and Borsotti et al.'s work had 26 participants~\cite{borsotti2024neurodiversity}. In two papers, the number of participants was not clearly reported~\cite{sitbon2018engaging, kletenik2024awareness}. 
Participants were not always exclusively neurodivergent.
While some user studies specifically recruited only neurodiverse participants for their studies~\cite{begel2021remote, haynes2024neurodiverse}, other user studies recruited a diverse participant group~\cite{dixon2007comparative, sitbon2018engaging, kletenik2024awareness, borsotti2024neurodiversity, kirdani2024neurodivergent}. When participant groups were mixed and the number of participants was specified~\cite{dixon2007comparative, borsotti2024neurodiversity, kirdani2024neurodivergent}, the percentage of neurodivergent participants ranged from 50\% to 71.4\%. This higher incidence of neurodivergence than is observed in the general population can be explained by the recruitment methods employed by the research teams. Dixon's sample of students were recruited from an introductory computing course which explains the higher incidence rate and was small (50\%)~\cite{dixon2007comparative}. Whereas Borsotti et al.~\cite{borsotti2024neurodiversity} and Kirdani-Ryan and Ko~\cite{kirdani2024neurodivergent} broadly recruited participants for a lab study, resulting 69.2\% and 71\% of participants being neurodivergent, respectively, explaining the higher incidence rate of neurodivergence. One recruitment method comes from Assiter's~\cite{assiter2018experiences} experience report about teaching a course for neurodiverse students at a college tailored for students with learning differences\footnote{\href{https://www.landmark.edu/}{Landmark College - https://www.landmark.edu/}}. As a result, all of their participants were neurodivergent~\cite{assiter2018experiences}.

\subsection{RQ2: Research Findings \& Outcomes}
\label{sec:rq2}

\subsubsection{Research Goals}

Most papers had the stated goal of designing a more inclusive curriculum (n=10). For example, two of the papers evaluated the accessibility of pedagogical approaches, such as algorithm visualization~\cite{dixon2007comparative} and Parsons problems~\cite{haynes2024neurodiverse}. The majority of these papers (n = 6), were dedicated to developing inclusive course curricula for neurodivergent students.
The topics of the curricula varied widely, from specific courses such as formal logic and ethics in computer science to broadly addressing issues across electrical and computing engineering, computer science, and computer science curricula.~\cite{ross2019supporting, martinez2011mhy, assiter2018experiences, pilotte2016autism, stuurman2019autism, egan2005students}. The final two papers provided guidelines developing accessible assessment techniques for autistic students~\cite{sharmin2024towards} and a course to raise awareness about the unique accessibility needs of neurodivergent software users~\cite{kletenik2024awareness}.

The remaining papers (n=4) were about diverse topics, such as developing communication skills for autistic students~\cite{begel2021remote}, engaging students in participatory design sessions with people who have intellectual disabilities~\cite{sitbon2018engaging}, identifying challenges faced by neurodivergent students~\cite{borsotti2024neurodiversity}, and investigating how different kinds of neurodivergence may be validated in computing spaces~\cite{kirdani2024neurodivergent}. 

\subsubsection{Creating Inclusive Curricula}
The primary findings for most papers (n=10) were recommendations to make curricula more inclusive. These recommendations were often motivated by specific neurodivergent populations, such as autistic students, which was the most common type that we observed~\cite{pilotte2016autism, stuurman2019autism, egan2005students, sharmin2024towards}. For example, autistic students tend to think very literally, which may cause them to struggle with ill-defined project requirements~\cite{egan2005students, sharmin2024towards, stuurman2019autism}. They also benefit from routines and may struggle with unexpected changes~\cite{pilotte2016autism, egan2005students}. Researchers recommended providing students with explicit instructions and rubrics for assignments~\cite{pilotte2016autism, stuurman2019autism, egan2005students, sharmin2024towards}, avoiding the use of ambiguous language, and using accessible visual representations of concepts~\cite{egan2005students, sharmin2024towards}.


Autistic students may also face issues with executive function, which can affect students' time management and concentration skills~\cite{stuurman2019autism, egan2005students, sharmin2024towards}. Providing alternative resources, assignments, and assessments was a common strategy to address challenges with executive function~\cite{sharmin2024towards, egan2005students}. 
Some specific strategies include recording the lectures~\cite{sharmin2024towards} and providing lecture notes~\cite{egan2005students}. 
These approaches were also used with dyslexic students. For example, materials were explicitly designed to help dyslexic students interpret the symbols used in formal logic courses~\cite{martinez2011mhy}.

In the discussion of one paper, the authors raised concerns that an increased emphasis on active learning strategies might have a negative impact on students who experience difficulties with social interactions, such as autistic students~\cite{pilotte2016autism}. Although other papers did not mention active learning, two papers recommended alternatives to group work to support autistic students~\cite{sharmin2024towards, stuurman2019autism}.
Additionally, Kletenik et al.~\cite{kletenik2024awareness} provided guidelines for including neurodiversity-centered accessibility education in computing courses, based on the access needs of autistic, dyslexic, and ADHD people. 
Ross~\cite{ross2019supporting} and Sharmin et al.~\cite{sharmin2024towards} highlighted the Universal Design for Learning (UDL)~\cite{rose2013universal} framework as a basis for designing inclusive courses~\cite{ross2019supporting, sharmin2024towards}. UDL has the potential to benefit all students, especially for students who may not have a formal diagnosis or wish not to self-disclose their neurodivergent status~\cite{sharmin2024towards}.

\subsubsection{Other Research Outcomes}

Developing a greater awareness of the needs of neurodivergent students was explicitly highlighted by 5 of the papers~\cite{borsotti2024neurodiversity, pilotte2016autism, kletenik2024awareness, egan2005students, martinez2011mhy}. 
Borsotti et al.~\cite{borsotti2024neurodiversity} highlighted the access needs of neurodivergent students and how to address and prevent barriers to access. They suggested ``access grafting''~\cite{borsotti2024neurodiversity} as a way to improve access through organizational change. Access grafting involves developing accessibility literacy and supporting neurodivergent-led interventions to improve access~\cite{borsotti2024neurodiversity}.

In addition to developing inclusive curricula and improving awareness, we identified some other key findings. Begel et al.~\cite{begel2021remote} studied how informal education, in the form of a video game coding camp, can be used to develop skills for incoming college students. Their camp improved the communication skills of their participants, and students exhibited early reflection skills that serve as a basis for classes to come. Sitbon~\cite{sitbon2018engaging} found that IT students were able to successfully facilitate co-design sessions, collaborating with people with intellectual disabilities with little training, largely due to strong feelings of reciprocity~\cite{duysburgh2015reciprocity} between them. Lastly, Kirdani-Ryan and Ko~\cite{kirdani2024neurodivergent} investigated how different neurotypes (e.g., kinds of neurodiversity) are encouraged or discouraged in computing spaces and found that some neurotypes (e.g., special interests, hyper-fixation, and organization) are encouraged, while some are discouraged (e.g., alternative career aspirations, non-computing-related special interests). However, they also found that computing spaces can be a place of refuge for those who identify as neurodivergent, despite some neurotypes being less validated than others.

\section{Discussion}


Our results show a growing focus on broadening computing to better support neurodivergent students. However, our review resulted in only 14 papers, suggesting that researchers still lack a thorough understanding of neurodivergent students' experiences and few best practices exist for effectively supporting these students. Our findings focus on two primary aspects:  1) the unique challenges faced by neurodivergent students and 2) the emerging guidelines for researchers and practitioners to better support these students.

\subsection{Challenges for Neurodivergent Students}

Our results suggest that neurodivergent computing students can experience challenges with collaboration, executive function, and visual or cognitive barriers to processing code. 
These findings align with broader research about neurodiversity in post-secondary education~\cite{clouder2020neurodiversity, zolyomi2018values}. This gives researchers and practitioners in CE an opportunity to explore more specific issues regarding the development of accessible and equitable curricula.
For example, the challenges neurodivergent students can face when collaborating are concerning given how enthusiastically many educators have embraced active learning~\cite{mccartney2017folk}. Although active learning provides learning benefits, and opportunities to develop social skills, our results suggest that it may inadvertently introduce barriers for neurodivergent students who may be less comfortable working in groups~\cite{pilotte2016autism, salvatore2024not} or struggle with executive function (i.e., experience difficulties with self-regulation, time management, and planning)~\cite{wischnewsky2023empowering}. 
Autistic students often benefit from predictable and well-structured environments, but they can struggle with the social dynamics and pace of active learning~\cite{gin2020active, pilotte2016autism}. ADHD students may find the flexibility of active learning stimulating, but can become overwhelmed without sufficient support to maintain their attention and manage distractions~\cite{pfeifer2023wish}. These observations show a potential tension between how active learning might impact autistic students versus other neurodivergent groups~\cite{butcher2024neurodivergent}.



The accessibility of course materials can be a barrier for students, especially for dyslexic students~\cite{dixon2007comparative, martinez2011mhy}. Verbal programming or multimodal learning tools may better support comprehension for all students compared to traditional text-based modalities. In the future, researchers can focus on developing new pedagogies and modalities for programming~\cite{mountapmbeme2022addressing, fuertes2018visual}.


\subsection{Accessibility Education \& Awareness}

There is a growing emphasis on integrating accessibility education. Examples from our review included developing curricula recommendations for teaching software accessibility~\cite{kletenik2024awareness} and increasing the awareness of access needs~\cite{borsotti2024neurodiversity}. Researchers are investigating ways to broadly integrate accessibility, not specifically related to neurodiversity, into specific courses~\cite{elglaly2024beyond, zhao2020comparison, shinohara2018teaches, baker2020systematic}. While this is an important endeavor, another approach we observed is to implement cross-curricular initiatives~\cite{shinohara2018teaches} that may result in comprehensive solutions rather than short-term fixes. 

A key issue raised in some of the papers we reviewed was a lack of awareness~\cite{martinez2011mhy, egan2005students, borsotti2024neurodiversity, pilotte2016autism} and institutional support~\cite{borsotti2024neurodiversity, kirdani2024neurodivergent} for neurodiverse students. These findings are not unique to computing education. Researchers and practitioners have identified many instances of awareness issues and institutional barriers neurodiverse students face, such as ineffective support services~\cite{evans2023autism, butcher2024neurodivergent}. Researchers and practitioners in CE can leverage this prior work to develop better awareness of students' access needs and support systems which do not place additional burden the students they are attempting to serve~\cite{clouder2020neurodiversity, evans2023autism}.

\subsection{Moving Forward}

Finally, our results show a missing aspect of designing effective educational environments---the active participation of neurodivergent students themselves. Echoing the ``nothing about us without us'' movement~\cite{charlton1998nothing}, researchers should engage more directly with neurodivergent students, co-designing pedagogical approaches that meet their needs. Only one of the studies that we analyzed used a participatory design methodology, which demonstrates the multiple missed opportunities to include neurodivergent voices in the creation of educational interventions. This gap points to a underlying issue in computing education research, where solutions are often designed without sufficient input from the students themselves.


There has been a tendency to treat disability groups as a homogeneous category~\cite{swain2008disability}. However, disabilities often intersect with other aspects of a person's identity, such as race, gender, and other disabilities~\cite{artiles2013untangling}. These intersections create experiences that cannot be understood by treating disability as a single, isolated factor. It is important to interpret and apply the findings of our work with caution. This study acts as a guide toward moving forward rather than a conclusive set of best practices.

\section{Limitations \& Future Work} 

Our search of the literature resulted in a small data set. We did not conduct snowballing, which may have impacted the size of our dataset. This is an emerging research area, though, and literature reviews with small sample sizes~\cite{martin2022intelligent} have contained meaningful insights. 
Despite limitations, this review identifies some emerging trends and future research directions for research about neurodiversity in computing education. Finally, we did not include papers about K-12 CE because we recognize that post-secondary education provides a unique context, and the needs of college and K-12 students are drastically different.

\section{Conclusion} 
We conducted a systematic literature to investigate what research has been done regarding neurodivergent students in post-secondary CE. Through our analysis of 14 selected papers, we found that interview studies and position papers have been the most prevalent in this research area. We also found that there were inconsistencies in reporting study details. We identify pathways for future research, including more empirical studies which directly involve neurodivergent students and research into active learning and neurodiversity-inclusive accessibility education.

\bibliographystyle{ACM-Reference-Format}
\balance
\bibliography{sample-base}


\begin{thebibliography}{61}


\ifx \showCODEN    \undefined \def \showCODEN     #1{\unskip}     \fi
\ifx \showDOI      \undefined \def \showDOI       #1{#1}\fi
\ifx \showISBNx    \undefined \def \showISBNx     #1{\unskip}     \fi
\ifx \showISBNxiii \undefined \def \showISBNxiii  #1{\unskip}     \fi
\ifx \showISSN     \undefined \def \showISSN      #1{\unskip}     \fi
\ifx \showLCCN     \undefined \def \showLCCN      #1{\unskip}     \fi
\ifx \shownote     \undefined \def \shownote      #1{#1}          \fi
\ifx \showarticletitle \undefined \def \showarticletitle #1{#1}   \fi
\ifx \showURL      \undefined \def \showURL       {\relax}        \fi
\providecommand\bibfield[2]{#2}
\providecommand\bibinfo[2]{#2}
\providecommand\natexlab[1]{#1}
\providecommand\showeprint[2][]{arXiv:#2}

\bibitem[Andresen et~al\mbox{.}(2001)]%
        {andresen2001proxy}
\bibfield{author}{\bibinfo{person}{Elena~M Andresen}, \bibinfo{person}{Victoria~J Vahle}, {and} \bibinfo{person}{Donald Lollar}.} \bibinfo{year}{2001}\natexlab{}.
\newblock \showarticletitle{Proxy reliability: health-related quality of life (HRQoL) measures for people with disability}.
\newblock \bibinfo{journal}{\emph{Quality of Life Research}}  \bibinfo{volume}{10} (\bibinfo{year}{2001}).
\newblock


\bibitem[Apiola et~al\mbox{.}(2023)]%
        {apiola2023past}
\bibfield{author}{\bibinfo{person}{Mikko Apiola}, \bibinfo{person}{Sonsoles L{\'o}pez-Pernas}, {and} \bibinfo{person}{Mohammed Saqr}.} \bibinfo{year}{2023}\natexlab{}.
\newblock \bibinfo{booktitle}{\emph{Past, Present and Future of Computing Education Research: A Global Perspective}}.
\newblock \bibinfo{publisher}{Springer Nature}.
\newblock


\bibitem[Artiles(2013)]%
        {artiles2013untangling}
\bibfield{author}{\bibinfo{person}{Alfredo~J Artiles}.} \bibinfo{year}{2013}\natexlab{}.
\newblock \showarticletitle{Untangling the racialization of disabilities: An intersectionality critique across disability models}.
\newblock \bibinfo{journal}{\emph{Du Bois Review: Social Science Research on Race}} (\bibinfo{year}{2013}).
\newblock


\bibitem[Assiter(2018)]%
        {assiter2018experiences}
\bibfield{author}{\bibinfo{person}{Karina Assiter}.} \bibinfo{year}{2018}\natexlab{}.
\newblock \showarticletitle{Experiences teaching a social and ethical aspects of computer science course to students with learning differences}.
\newblock \bibinfo{journal}{\emph{Journal of Computing Sciences in Colleges}} \bibinfo{volume}{34}, \bibinfo{number}{2} (\bibinfo{year}{2018}).
\newblock


\bibitem[Baker et~al\mbox{.}(2020)]%
        {baker2020systematic}
\bibfield{author}{\bibinfo{person}{Catherine~M Baker}, \bibinfo{person}{Yasmine~N El-Glaly}, {and} \bibinfo{person}{Kristen Shinohara}.} \bibinfo{year}{2020}\natexlab{}.
\newblock \showarticletitle{A systematic analysis of accessibility in computing education research}. In \bibinfo{booktitle}{\emph{Proceedings of the 51st ACM Technical Symposium on Computer Science Education}}.
\newblock


\bibitem[Begel et~al\mbox{.}(2021)]%
        {begel2021remote}
\bibfield{author}{\bibinfo{person}{Andrew Begel}, \bibinfo{person}{James Dominic}, \bibinfo{person}{Conner Phillis}, \bibinfo{person}{Thomas Beeson}, {and} \bibinfo{person}{Paige Rodeghero}.} \bibinfo{year}{2021}\natexlab{}.
\newblock \showarticletitle{How a Remote Video Game Coding Camp Improved Autistic College Students' Self-Efficacy in Communication}. In \bibinfo{booktitle}{\emph{Proceedings of the 52nd ACM Technical Symposium on Computer Science Education}}.
\newblock


\bibitem[Blaser and Ladner(2020)]%
        {blaser2020why}
\bibfield{author}{\bibinfo{person}{Brianna Blaser} {and} \bibinfo{person}{Richard~E. Ladner}.} \bibinfo{year}{2020}\natexlab{}.
\newblock \showarticletitle{Why is Data on Disability so Hard to Collect and Understand?}. In \bibinfo{booktitle}{\emph{2020 Research on Equity and Sustained Participation in Engineering, Computing, and Technology}}.
\newblock


\bibitem[Blaser et~al\mbox{.}(2024)]%
        {blaser2024accessibility}
\bibfield{author}{\bibinfo{person}{Brianna Blaser}, \bibinfo{person}{Richard~E. Ladner}, \bibinfo{person}{Bryan Twarek}, \bibinfo{person}{Andreas Stefik}, {and} \bibinfo{person}{Hannah Stabler}.} \bibinfo{year}{2024}\natexlab{}.
\newblock \showarticletitle{Accessibility and Disability in PreK-12 CS: Results from a Landscape Survey of Teachers}. In \bibinfo{booktitle}{\emph{Proceedings of the 2024 on RESPECT Annual Conference}}.
\newblock
\showISBNx{9798400706264}


\bibitem[Borsotti et~al\mbox{.}(2024)]%
        {borsotti2024neurodiversity}
\bibfield{author}{\bibinfo{person}{Valeria Borsotti}, \bibinfo{person}{Andrew Begel}, {and} \bibinfo{person}{Pernille Bj\o{}rn}.} \bibinfo{year}{2024}\natexlab{}.
\newblock \showarticletitle{Neurodiversity and the Accessible University: Exploring Organizational Barriers, Access Labor and Opportunities for Change}.
\newblock \bibinfo{journal}{\emph{Proc. ACM Hum.-Comput. Interact.}} \bibinfo{volume}{8}, \bibinfo{number}{CSCW1} (\bibinfo{year}{2024}).
\newblock


\bibitem[Butcher and Lane(2024)]%
        {butcher2024neurodivergent}
\bibfield{author}{\bibinfo{person}{Luke Butcher} {and} \bibinfo{person}{Stevie Lane}.} \bibinfo{year}{2024}\natexlab{}.
\newblock \showarticletitle{Neurodivergent (Autism and ADHD) student experiences of access and inclusion in higher education: an ecological systems theory perspective}.
\newblock \bibinfo{journal}{\emph{Higher Education}} (\bibinfo{year}{2024}).
\newblock


\bibitem[Charlton(1998)]%
        {charlton1998nothing}
\bibfield{author}{\bibinfo{person}{James~I Charlton}.} \bibinfo{year}{1998}\natexlab{}.
\newblock \bibinfo{booktitle}{\emph{Nothing about us without us: Disability oppression and empowerment}}.
\newblock \bibinfo{publisher}{Univ of California Press}.
\newblock


\bibitem[Clouder et~al\mbox{.}(2020)]%
        {clouder2020neurodiversity}
\bibfield{author}{\bibinfo{person}{Lynn Clouder}, \bibinfo{person}{Mehmet Karakus}, \bibinfo{person}{Alessia Cinotti}, \bibinfo{person}{Mar{\'\i}a~Virginia Ferreyra}, \bibinfo{person}{Genoveva~Amador Fierros}, {and} \bibinfo{person}{Patricia Rojo}.} \bibinfo{year}{2020}\natexlab{}.
\newblock \showarticletitle{Neurodiversity in higher education: a narrative synthesis}.
\newblock \bibinfo{journal}{\emph{Higher Education}} \bibinfo{volume}{80}, \bibinfo{number}{4} (\bibinfo{year}{2020}), \bibinfo{pages}{757--778}.
\newblock


\bibitem[Cole and Espinoza(2008)]%
        {cole2008examining}
\bibfield{author}{\bibinfo{person}{Darnell Cole} {and} \bibinfo{person}{Araceli Espinoza}.} \bibinfo{year}{2008}\natexlab{}.
\newblock \showarticletitle{Examining the academic success of Latino students in science technology engineering and mathematics (STEM) majors}.
\newblock \bibinfo{journal}{\emph{Journal of College Student Development}} \bibinfo{volume}{49}, \bibinfo{number}{4} (\bibinfo{year}{2008}).
\newblock


\bibitem[Cotler et~al\mbox{.}(2024)]%
        {cotler2024enhancing}
\bibfield{author}{\bibinfo{person}{Jami Cotler}, \bibinfo{person}{Eszter Kiss}, \bibinfo{person}{Dmitry Burshteyn}, \bibinfo{person}{Megan Hale}, \bibinfo{person}{Amani Walker}, {and} \bibinfo{person}{John Slyer}.} \bibinfo{year}{2024}\natexlab{}.
\newblock \showarticletitle{Enhancing Empathy and Inclusivity in Computer Science Education: An Empirical Study on Accessibility Interventions for Undergraduate Students}.
\newblock \bibinfo{journal}{\emph{J. Comput. Sci. Coll.}} \bibinfo{volume}{39}, \bibinfo{number}{8} (\bibinfo{date}{May} \bibinfo{year}{2024}).
\newblock
\showISSN{1937-4771}


\bibitem[Dixon(2007)]%
        {dixon2007comparative}
\bibfield{author}{\bibinfo{person}{Mark Dixon}.} \bibinfo{year}{2007}\natexlab{}.
\newblock \showarticletitle{Comparative study of disabled vs. non-disabled evaluators in user-testing: dyslexia and first year students learning computer programming}. In \bibinfo{booktitle}{\emph{Universal Acess in Human Computer Interaction. Coping with Diversity: 4th International Conference on Universal Access in Human-Computer Interaction}}.
\newblock


\bibitem[Duysburgh and Slegers(2015)]%
        {duysburgh2015reciprocity}
\bibfield{author}{\bibinfo{person}{Pieter Duysburgh} {and} \bibinfo{person}{Karin Slegers}.} \bibinfo{year}{2015}\natexlab{}.
\newblock \showarticletitle{Reciprocity in rapid ethnography: Giving back by making the small things count}. In \bibinfo{booktitle}{\emph{Human-Computer Interaction--INTERACT 2015: 15th IFIP TC 13 International Conference}}. Springer.
\newblock


\bibitem[Egan(2005)]%
        {egan2005students}
\bibfield{author}{\bibinfo{person}{Mary Anne~L Egan}.} \bibinfo{year}{2005}\natexlab{}.
\newblock \showarticletitle{Students with Asperger's syndrome in the CS classroom}. In \bibinfo{booktitle}{\emph{Proceedings of the 36th SIGCSE technical symposium on computer science education}}.
\newblock


\bibitem[Elglaly et~al\mbox{.}(2024)]%
        {elglaly2024beyond}
\bibfield{author}{\bibinfo{person}{Yasmine~N Elglaly}, \bibinfo{person}{Catherine~M Baker}, \bibinfo{person}{Anne~Spencer Ross}, {and} \bibinfo{person}{Kristen Shinohara}.} \bibinfo{year}{2024}\natexlab{}.
\newblock \showarticletitle{Beyond HCI: The Need for Accessibility Across the CS Curriculum}. In \bibinfo{booktitle}{\emph{Proceedings of the 55th ACM Technical Symposium on Computer Science Education}}.
\newblock


\bibitem[Ericson et~al\mbox{.}(2022)]%
        {ericson2022parsons}
\bibfield{author}{\bibinfo{person}{Barbara~J Ericson}, \bibinfo{person}{Paul Denny}, \bibinfo{person}{James Prather}, {et~al\mbox{.}}} \bibinfo{year}{2022}\natexlab{}.
\newblock \showarticletitle{Parsons problems and beyond: Systematic literature review and empirical study designs}.
\newblock \bibinfo{journal}{\emph{Proceedings of the 2022 Working Group Reports on Innovation and Technology in Computer Science Education}} (\bibinfo{year}{2022}).
\newblock


\bibitem[Evans et~al\mbox{.}(2023)]%
        {evans2023autism}
\bibfield{author}{\bibinfo{person}{Dena Evans}, \bibinfo{person}{Matthew Granson}, \bibinfo{person}{David Langford}, {and} \bibinfo{person}{Sophie Hirsch}.} \bibinfo{year}{2023}\natexlab{}.
\newblock \showarticletitle{Autism spectrum disorder: reconceptualising support for neurodiverse students in higher education}.
\newblock \bibinfo{journal}{\emph{Journal of Higher Education Policy and Management}} \bibinfo{volume}{45}, \bibinfo{number}{2} (\bibinfo{year}{2023}), \bibinfo{pages}{243--257}.
\newblock


\bibitem[Fuertes et~al\mbox{.}(2018)]%
        {fuertes2018visual}
\bibfield{author}{\bibinfo{person}{Jose~L Fuertes}, \bibinfo{person}{Luis~F Gonzalez}, {and} \bibinfo{person}{Loic Martinez}.} \bibinfo{year}{2018}\natexlab{}.
\newblock \showarticletitle{Visual programming languages for programmers with dyslexia: An experiment}. In \bibinfo{booktitle}{\emph{2018 IEEE 14th International Conference on e-Science (e-Science)}}. IEEE.
\newblock


\bibitem[Gin et~al\mbox{.}(2020)]%
        {gin2020active}
\bibfield{author}{\bibinfo{person}{Logan~E Gin}, \bibinfo{person}{Frank~A Guerrero}, \bibinfo{person}{Katelyn~M Cooper}, {and} \bibinfo{person}{Sara~E Brownell}.} \bibinfo{year}{2020}\natexlab{}.
\newblock \showarticletitle{Is active learning accessible? Exploring the process of providing accommodations to students with disabilities}.
\newblock \bibinfo{journal}{\emph{CBE—Life Sciences Education}} (\bibinfo{year}{2020}).
\newblock


\bibitem[Haynes-Magyar(2024)]%
        {haynes2024neurodiverse}
\bibfield{author}{\bibinfo{person}{Carl Haynes-Magyar}.} \bibinfo{year}{2024}\natexlab{}.
\newblock \showarticletitle{Neurodiverse Programmers and the Accessibility of Parsons Problems: An Exploratory Multiple-Case Study}. In \bibinfo{booktitle}{\emph{Proceedings of the 55th ACM Technical Symposium on Computer Science Education V. 1}}.
\newblock


\bibitem[Israel et~al\mbox{.}(2015)]%
        {israel2015empowering}
\bibfield{author}{\bibinfo{person}{Maya Israel}, \bibinfo{person}{Quentin~M Wherfel}, \bibinfo{person}{Jamie Pearson}, \bibinfo{person}{Saadeddine Shehab}, {and} \bibinfo{person}{Tanya Tapia}.} \bibinfo{year}{2015}\natexlab{}.
\newblock \showarticletitle{Empowering K--12 students with disabilities to learn computational thinking and computer programming}.
\newblock \bibinfo{journal}{\emph{TEACHING Exceptional Children}} (\bibinfo{year}{2015}).
\newblock


\bibitem[Kirdani-Ryan and Ko(2024)]%
        {kirdani2024neurodivergent}
\bibfield{author}{\bibinfo{person}{Mara Kirdani-Ryan} {and} \bibinfo{person}{Amy~J Ko}.} \bibinfo{year}{2024}\natexlab{}.
\newblock \showarticletitle{Neurodivergent Legitimacy in Computing Spaces}.
\newblock \bibinfo{journal}{\emph{ACM Transactions on Computing Education}} \bibinfo{volume}{24}, \bibinfo{number}{4} (\bibinfo{year}{2024}).
\newblock


\bibitem[Kletenik et~al\mbox{.}(2024)]%
        {kletenik2024awareness}
\bibfield{author}{\bibinfo{person}{Devorah Kletenik}, \bibinfo{person}{Rachel Minkowitz}, \bibinfo{person}{Aleksandra Peric}, \bibinfo{person}{Mehmet Sahin}, {and} \bibinfo{person}{Rachel~F Adler}.} \bibinfo{year}{2024}\natexlab{}.
\newblock \showarticletitle{From Awareness to Action: Teaching Software Accessibility for Neurodiverse Users}. In \bibinfo{booktitle}{\emph{Proceedings of the 55th ACM Technical Symposium on Computer Science Education}}.
\newblock


\bibitem[Koushik and Kane(2019)]%
        {koushik2019broadens}
\bibfield{author}{\bibinfo{person}{Varsha Koushik} {and} \bibinfo{person}{Shaun~K Kane}.} \bibinfo{year}{2019}\natexlab{}.
\newblock \showarticletitle{" It Broadens My Mind" Empowering People with Cognitive Disabilities through Computing Education}. In \bibinfo{booktitle}{\emph{Proceedings of the 2019 CHI conference on human factors in computing systems}}.
\newblock


\bibitem[London et~al\mbox{.}(2020)]%
        {london2020systematic}
\bibfield{author}{\bibinfo{person}{Jeremi~S London}, \bibinfo{person}{Walter~C Lee}, \bibinfo{person}{Canek Phillips}, \bibinfo{person}{Amy~S Van~Epps}, {and} \bibinfo{person}{Bevlee~A Watford}.} \bibinfo{year}{2020}\natexlab{}.
\newblock \showarticletitle{A systematic mapping of scholarship on broadening participation of African Americans in engineering and computer science}.
\newblock \bibinfo{journal}{\emph{Journal of Women and Minorities in Science and Engineering}} \bibinfo{volume}{26}, \bibinfo{number}{3} (\bibinfo{year}{2020}).
\newblock


\bibitem[Lor{\aa}s et~al\mbox{.}(2021)]%
        {loraas2021study}
\bibfield{author}{\bibinfo{person}{Madeleine Lor{\aa}s}, \bibinfo{person}{Guttorm Sindre}, \bibinfo{person}{Hallvard Tr{\ae}tteberg}, {and} \bibinfo{person}{Trond Aalberg}.} \bibinfo{year}{2021}\natexlab{}.
\newblock \showarticletitle{Study behavior in computing education—a systematic literature review}.
\newblock \bibinfo{journal}{\emph{ACM Transactions on Computing Education (TOCE)}} \bibinfo{volume}{22}, \bibinfo{number}{1} (\bibinfo{year}{2021}).
\newblock


\bibitem[Luchs(2021)]%
        {luchs2021considering}
\bibfield{author}{\bibinfo{person}{Christopher Luchs}.} \bibinfo{year}{2021}\natexlab{}.
\newblock \showarticletitle{Considering Neurodiversity in Learning Design and Technology}.
\newblock \bibinfo{journal}{\emph{TechTrends}} \bibinfo{volume}{65}, \bibinfo{number}{6} (\bibinfo{year}{2021}).
\newblock


\bibitem[Lutz and Bowers(2005)]%
        {lutz2005disability}
\bibfield{author}{\bibinfo{person}{Barbara~J Lutz} {and} \bibinfo{person}{Barbara~J Bowers}.} \bibinfo{year}{2005}\natexlab{}.
\newblock \showarticletitle{Disability in everyday life}.
\newblock \bibinfo{journal}{\emph{Qualitative Health Research}} \bibinfo{volume}{15}, \bibinfo{number}{8} (\bibinfo{year}{2005}).
\newblock


\bibitem[Mack et~al\mbox{.}(2021)]%
        {mack2021we}
\bibfield{author}{\bibinfo{person}{Kelly Mack}, \bibinfo{person}{Emma McDonnell}, \bibinfo{person}{Dhruv Jain}, \bibinfo{person}{Lucy Lu~Wang}, \bibinfo{person}{Jon E.~Froehlich}, {and} \bibinfo{person}{Leah Findlater}.} \bibinfo{year}{2021}\natexlab{}.
\newblock \showarticletitle{What do we mean by “accessibility research”? A literature survey of accessibility papers in CHI and ASSETS from 1994 to 2019}. In \bibinfo{booktitle}{\emph{Proceedings of the 2021 CHI Conference on Human Factors in Computing Systems}}.
\newblock


\bibitem[Martin et~al\mbox{.}(2022)]%
        {martin2022intelligent}
\bibfield{author}{\bibinfo{person}{Alexia~Charis Martin}, \bibinfo{person}{Kimberly~Michelle Ying}, \bibinfo{person}{Fernando~J Rodr{\'\i}guez}, {et~al\mbox{.}}} \bibinfo{year}{2022}\natexlab{}.
\newblock \showarticletitle{Intelligent Support for All? A Literature Review of the (In) equitable Design \& Evaluation of Adaptive Pedagogical Systems for CS Education}. In \bibinfo{booktitle}{\emph{Proceedings of the 53rd ACM Technical Symposium on Computer Science Education-Volume 1}}.
\newblock


\bibitem[Mart{\'\i}nez~Nava(2011)]%
        {martinez2011mhy}
\bibfield{author}{\bibinfo{person}{X{\'o}chitl Mart{\'\i}nez~Nava}.} \bibinfo{year}{2011}\natexlab{}.
\newblock \showarticletitle{Mhy bib i fail logic? Dyslexia in the teaching of logic}. In \bibinfo{booktitle}{\emph{International Congress on Tools for Teaching Logic}}. Springer.
\newblock


\bibitem[McCartney et~al\mbox{.}(2017)]%
        {mccartney2017folk}
\bibfield{author}{\bibinfo{person}{Robert McCartney}, \bibinfo{person}{Jonas Boustedt}, \bibinfo{person}{Anna Eckerdal}, \bibinfo{person}{Kate Sanders}, {and} \bibinfo{person}{Carol Zander}.} \bibinfo{year}{2017}\natexlab{}.
\newblock \showarticletitle{Folk Pedagogy and the Geek Gene: Geekiness Quotient}. In \bibinfo{booktitle}{\emph{Proceedings of the 2017 ACM SIGCSE Technical Symposium on Computer Science Education}}.
\newblock
\showISBNx{9781450346986}


\bibitem[Menier et~al\mbox{.}(2021)]%
        {menier2021broadening}
\bibfield{author}{\bibinfo{person}{Amanda Menier}, \bibinfo{person}{Rebecca Zarch}, {and} \bibinfo{person}{Stacey Sexton}.} \bibinfo{year}{2021}\natexlab{}.
\newblock \showarticletitle{Broadening Gender in Computing for Transgender and Nonbinary Learners}. In \bibinfo{booktitle}{\emph{2021 Conference on Research in Equitable and Sustained Participation in Engineering, Computing, and Technology (RESPECT)}}.
\newblock


\bibitem[Michell et~al\mbox{.}(2017)]%
        {michell2017broadening}
\bibfield{author}{\bibinfo{person}{Dee Michell}, \bibinfo{person}{Anna Szorenyi}, \bibinfo{person}{Katrina Falkner}, {and} \bibinfo{person}{Claudia Szabo}.} \bibinfo{year}{2017}\natexlab{}.
\newblock \showarticletitle{Broadening participation not border protection: How universities can support women in computer science}.
\newblock \bibinfo{journal}{\emph{Journal of Higher Education Policy and Management}} (\bibinfo{year}{2017}).
\newblock


\bibitem[Mountapmbeme et~al\mbox{.}(2022)]%
        {mountapmbeme2022addressing}
\bibfield{author}{\bibinfo{person}{Aboubakar Mountapmbeme}, \bibinfo{person}{Obianuju Okafor}, {and} \bibinfo{person}{Stephanie Ludi}.} \bibinfo{year}{2022}\natexlab{}.
\newblock \showarticletitle{Addressing accessibility barriers in programming for people with visual impairments: A literature review}.
\newblock \bibinfo{journal}{\emph{ACM Transactions on Accessible Computing}} (\bibinfo{year}{2022}).
\newblock


\bibitem[Page et~al\mbox{.}(2021)]%
        {page2021prisma}
\bibfield{author}{\bibinfo{person}{Matthew~J Page}, \bibinfo{person}{Joanne~E McKenzie}, {et~al\mbox{.}}} \bibinfo{year}{2021}\natexlab{}.
\newblock \showarticletitle{The PRISMA 2020 statement: an updated guideline for reporting systematic reviews}.
\newblock \bibinfo{journal}{\emph{BMJ}}  \bibinfo{volume}{372} (\bibinfo{year}{2021}).
\newblock


\bibitem[Pfeifer et~al\mbox{.}(2023)]%
        {pfeifer2023wish}
\bibfield{author}{\bibinfo{person}{Mariel~A Pfeifer}, \bibinfo{person}{Julio~J Cordero}, {and} \bibinfo{person}{Julie~Dangremond Stanton}.} \bibinfo{year}{2023}\natexlab{}.
\newblock \showarticletitle{What I wish my instructor knew: How active learning influences the classroom experiences and self-advocacy of STEM majors with ADHD and specific learning disabilities}.
\newblock \bibinfo{journal}{\emph{CBE—Life Sciences Education}} (\bibinfo{year}{2023}).
\newblock


\bibitem[Pilotte and Bairaktarova(2016)]%
        {pilotte2016autism}
\bibfield{author}{\bibinfo{person}{Mary Pilotte} {and} \bibinfo{person}{Diana Bairaktarova}.} \bibinfo{year}{2016}\natexlab{}.
\newblock \showarticletitle{Autism spectrum disorder and engineering education-needs and considerations}. In \bibinfo{booktitle}{\emph{2016 IEEE Frontiers in Education Conference (FIE)}}. IEEE.
\newblock


\bibitem[Regier et~al\mbox{.}(2013)]%
        {regier2013dsm}
\bibfield{author}{\bibinfo{person}{Darrel~A Regier}, \bibinfo{person}{Emily~A Kuhl}, {and} \bibinfo{person}{David~J Kupfer}.} \bibinfo{year}{2013}\natexlab{}.
\newblock \showarticletitle{The DSM-5: Classification and criteria changes}.
\newblock \bibinfo{journal}{\emph{World psychiatry}} \bibinfo{volume}{12}, \bibinfo{number}{2} (\bibinfo{year}{2013}).
\newblock


\bibitem[Rose et~al\mbox{.}(2013)]%
        {rose2013universal}
\bibfield{author}{\bibinfo{person}{David~H Rose}, \bibinfo{person}{Jenna~W Gravel}, {and} \bibinfo{person}{David~T Gordon}.} \bibinfo{year}{2013}\natexlab{}.
\newblock \showarticletitle{Universal design for learning}.
\newblock \bibinfo{journal}{\emph{The SAGE Handbook of Special Education: Two Volume Set}} (\bibinfo{year}{2013}).
\newblock


\bibitem[Ross(2019)]%
        {ross2019supporting}
\bibfield{author}{\bibinfo{person}{Sheila~R Ross}.} \bibinfo{year}{2019}\natexlab{}.
\newblock \showarticletitle{Supporting your neurodiverse student population with the Universal Design for Learning (UDL) framework}. In \bibinfo{booktitle}{\emph{2019 IEEE Frontiers in Education Conference (FIE)}}. IEEE.
\newblock


\bibitem[Runa et~al\mbox{.}(2023)]%
        {runa2023student}
\bibfield{author}{\bibinfo{person}{Shamima~Nasrin Runa}, \bibinfo{person}{Anna~Markella Antoniadi}, \bibinfo{person}{Brett~A. Becker}, {and} \bibinfo{person}{Catherine Mooney}.} \bibinfo{year}{2023}\natexlab{}.
\newblock \showarticletitle{Student Sense of Belonging: The Role of Gender Identity and Minoritisation in Computing and Other Sciences}. In \bibinfo{booktitle}{\emph{Proceedings of the 25th Australasian Computing Education Conference}} \emph{(\bibinfo{series}{ACE '23})}. \bibinfo{publisher}{ACM}.
\newblock
\showISBNx{9781450399418}


\bibitem[Salvatore et~al\mbox{.}(2024)]%
        {salvatore2024not}
\bibfield{author}{\bibinfo{person}{Sophia Salvatore}, \bibinfo{person}{Claudia White}, {and} \bibinfo{person}{Stephen Podowitz-Thomas}.} \bibinfo{year}{2024}\natexlab{}.
\newblock \showarticletitle{“Not a cookie cutter situation”: how neurodivergent students experience group work in their STEM courses}.
\newblock \bibinfo{journal}{\emph{International Journal of STEM Education}} \bibinfo{volume}{11}, \bibinfo{number}{1} (\bibinfo{year}{2024}), \bibinfo{pages}{47}.
\newblock


\bibitem[Santhanam and Wilson(2024)]%
        {santhanam2024comparison}
\bibfield{author}{\bibinfo{person}{Siva~priya Santhanam} {and} \bibinfo{person}{Kaitlyn Wilson}.} \bibinfo{year}{2024}\natexlab{}.
\newblock \showarticletitle{A Comparison of Autistic and Non-Autistic College Students' Perceived Challenges and Engagement in Self-Advocacy}.
\newblock \bibinfo{journal}{\emph{American Journal of Speech-Language Pathology}} \bibinfo{volume}{33}, \bibinfo{number}{3} (\bibinfo{year}{2024}).
\newblock


\bibitem[Scott et~al\mbox{.}(2017)]%
        {scott2017broadening}
\bibfield{author}{\bibinfo{person}{Allison Scott}, \bibinfo{person}{Alexis Martin}, \bibinfo{person}{Frieda McAlear}, {and} \bibinfo{person}{Sonia Koshy}.} \bibinfo{year}{2017}\natexlab{}.
\newblock \showarticletitle{Broadening Participation in Computing: Examining Experiences of Girls of Color}. In \bibinfo{booktitle}{\emph{Proceedings of the 2017 ACM Conference on Innovation and Technology in Computer Science Education}} \emph{(\bibinfo{series}{ITiCSE '17})}. \bibinfo{publisher}{ACM}.
\newblock
\showISBNx{9781450347044}


\bibitem[Sharif et~al\mbox{.}(2022)]%
        {sharif2022should}
\bibfield{author}{\bibinfo{person}{Ather Sharif}, \bibinfo{person}{Aedan~Liam McCall}, {and} \bibinfo{person}{Kianna~Roces Bolante}.} \bibinfo{year}{2022}\natexlab{}.
\newblock \showarticletitle{Should I Say “Disabled People” or “People with Disabilities”? Language Preferences of Disabled People Between Identity- and Person-First Language}. In \bibinfo{booktitle}{\emph{Proceedings of the 24th International ACM SIGACCESS Conference on Computers and Accessibility}}.
\newblock
\showISBNx{9781450392587}


\bibitem[Sharmin et~al\mbox{.}(2024)]%
        {sharmin2024towards}
\bibfield{author}{\bibinfo{person}{Moushumi Sharmin}, \bibinfo{person}{Jordan Archer}, \bibinfo{person}{Adrian Heffelman}, \bibinfo{person}{Eli Graves}, \bibinfo{person}{Yasmine~N Elglaly}, {and} \bibinfo{person}{Shameem Ahmed}.} \bibinfo{year}{2024}\natexlab{}.
\newblock \showarticletitle{Towards Understanding the Challenges, Needs, and Opportunities Pertaining to Assessment Techniques for Autistic College Students in Computing}. In \bibinfo{booktitle}{\emph{2024 IEEE 48th Annual Computers, Software, and Applications Conference (COMPSAC)}}. IEEE.
\newblock


\bibitem[Shehzad et~al\mbox{.}(2023)]%
        {shehzad2023literature}
\bibfield{author}{\bibinfo{person}{Umar Shehzad}, \bibinfo{person}{Mimi Recker}, {and} \bibinfo{person}{Jody Clarke-Midura}.} \bibinfo{year}{2023}\natexlab{}.
\newblock \showarticletitle{A literature review examining broadening participation in upper elementary CS education}. In \bibinfo{booktitle}{\emph{Proceedings of the 54th ACM Technical Symposium on Computer Science Education}}.
\newblock


\bibitem[Shinohara et~al\mbox{.}(2018)]%
        {shinohara2018teaches}
\bibfield{author}{\bibinfo{person}{Kristen Shinohara}, \bibinfo{person}{Saba Kawas}, \bibinfo{person}{Amy~J Ko}, {and} \bibinfo{person}{Richard~E Ladner}.} \bibinfo{year}{2018}\natexlab{}.
\newblock \showarticletitle{Who teaches accessibility? A survey of US computing faculty}. In \bibinfo{booktitle}{\emph{Proceedings of the 49th ACM Technical Symposium on Computer Science Education}}.
\newblock


\bibitem[Shinohara et~al\mbox{.}(2020)]%
        {shinohara2020access}
\bibfield{author}{\bibinfo{person}{Kristen Shinohara}, \bibinfo{person}{Michael McQuaid}, {and} \bibinfo{person}{Nayeri Jacobo}.} \bibinfo{year}{2020}\natexlab{}.
\newblock \showarticletitle{Access Differential and Inequitable Access: Inaccessibility for Doctoral Students in Computing}. In \bibinfo{booktitle}{\emph{Proceedings of the 22nd International ACM SIGACCESS Conference on Computers and Accessibility}} \emph{(\bibinfo{series}{ASSETS '20})}. \bibinfo{publisher}{ACM}, Article \bibinfo{articleno}{7}.
\newblock
\showISBNx{9781450371032}


\bibitem[Sitbon(2018)]%
        {sitbon2018engaging}
\bibfield{author}{\bibinfo{person}{Laurianne Sitbon}.} \bibinfo{year}{2018}\natexlab{}.
\newblock \showarticletitle{Engaging IT students in co-design with people with intellectual disability}. In \bibinfo{booktitle}{\emph{Extended Abstracts of the 2018 CHI Conference on Human Factors in Computing Systems}}.
\newblock


\bibitem[Stuurman et~al\mbox{.}(2019)]%
        {stuurman2019autism}
\bibfield{author}{\bibinfo{person}{Sylvia Stuurman}, \bibinfo{person}{Harrie~JM Passier}, \bibinfo{person}{Fr{\'e}d{\'e}rique Geven}, {and} \bibinfo{person}{Erik Barendsen}.} \bibinfo{year}{2019}\natexlab{}.
\newblock \showarticletitle{Autism: Implications for inclusive education with respect to software engineering}. In \bibinfo{booktitle}{\emph{Proceedings of the 8th Computer Science Education Research Conference}}.
\newblock


\bibitem[Swain and French(2008)]%
        {swain2008disability}
\bibfield{author}{\bibinfo{person}{John Swain} {and} \bibinfo{person}{Sally French}.} \bibinfo{year}{2008}\natexlab{}.
\newblock \bibinfo{booktitle}{\emph{Disability on equal terms}}.
\newblock \bibinfo{publisher}{Sage}.
\newblock


\bibitem[Van~Hees et~al\mbox{.}(2015)]%
        {van2015higher}
\bibfield{author}{\bibinfo{person}{Val{\'e}rie Van~Hees}, \bibinfo{person}{Tinneke Moyson}, {and} \bibinfo{person}{Herbert Roeyers}.} \bibinfo{year}{2015}\natexlab{}.
\newblock \showarticletitle{Higher education experiences of students with autism spectrum disorder: Challenges, benefits and support needs}.
\newblock \bibinfo{journal}{\emph{Journal of autism and developmental disorders}}  \bibinfo{volume}{45} (\bibinfo{year}{2015}).
\newblock


\bibitem[Wischnewsky(2023)]%
        {wischnewsky2023empowering}
\bibfield{author}{\bibinfo{person}{Lori~A Wischnewsky}.} \bibinfo{year}{2023}\natexlab{}.
\newblock \showarticletitle{Empowering Autistic College Students: Recommendations Based on a Review of the Literature and Existing Support Programs.}
\newblock \bibinfo{journal}{\emph{Journal of College Academic Support Programs}} (\bibinfo{year}{2023}).
\newblock


\bibitem[Wobbrock and Kientz(2016)]%
        {wobbrock2016research}
\bibfield{author}{\bibinfo{person}{Jacob~O Wobbrock} {and} \bibinfo{person}{Julie~A Kientz}.} \bibinfo{year}{2016}\natexlab{}.
\newblock \showarticletitle{Research contributions in human-computer interaction}.
\newblock \bibinfo{journal}{\emph{interactions}} \bibinfo{volume}{23}, \bibinfo{number}{3} (\bibinfo{year}{2016}).
\newblock


\bibitem[Zhao et~al\mbox{.}(2020)]%
        {zhao2020comparison}
\bibfield{author}{\bibinfo{person}{Qiwen Zhao}, \bibinfo{person}{Vaishnavi Mande}, \bibinfo{person}{Paula Conn}, {et~al\mbox{.}}} \bibinfo{year}{2020}\natexlab{}.
\newblock \showarticletitle{Comparison of methods for teaching accessibility in university computing courses}. In \bibinfo{booktitle}{\emph{Proceedings of the 22nd International ACM SIGACCESS Conference on Computers and Accessibility}}.
\newblock


\bibitem[Zolyomi et~al\mbox{.}(2018)]%
        {zolyomi2018values}
\bibfield{author}{\bibinfo{person}{Annuska Zolyomi}, \bibinfo{person}{Anne~Spencer Ross}, \bibinfo{person}{Arpita Bhattacharya}, \bibinfo{person}{Lauren Milne}, {and} \bibinfo{person}{Sean~A. Munson}.} \bibinfo{year}{2018}\natexlab{}.
\newblock \showarticletitle{Values, Identity, and Social Translucence: Neurodiverse Student Teams in Higher Education}. In \bibinfo{booktitle}{\emph{Proceedings of the 2018 CHI Conference on Human Factors in Computing Systems}}.
\newblock
\showISBNx{9781450356206}


\end{thebibliography}


\end{document}